\documentclass[useAMS, usenatbib, usegraphicx]{mn2e}
\usepackage{color}

\title[Mass segregation trends in SDSS galaxy groups]{Mass segregation
  trends in SDSS galaxy groups}

\author[I.D. Roberts et al]
       {Ian~D.~Roberts\thanks{E-mail: roberid@mcmaster.ca},
         Laura~C.~Parker, Gandhali~D.~Joshi and Fraser~A.~Evans \\
         Department of Physics and Astronomy, McMaster University,
         Hamilton ON L8S 4M1, Canada}

\begin{document}

\pagerange{000-000} \pubyear{0000}

\maketitle

\begin{abstract}
It has been shown that galaxy properties depend strongly on their host
environment. In order to understand the relevant physical processes
driving galaxy evolution it is important to study the observed
properties of galaxies in different environments. Mass
segregation in bound galaxy
structures is an
important indicator of evolutionary history and dynamical friction
timescales. Using group catalogues derived from the Sloan Digital Sky
Survey Data Release 7 (SDSS DR7) we
investigate mass segregation trends in galaxy groups at low
redshift. We investigate average galaxy stellar mass as a function of
group-centric radius and find evidence for weak mass segregation in
SDSS
groups. The
magnitude of the mass segregation depends on both galaxy stellar mass
limits and group halo
mass. We show that the inclusion of low mass galaxies tends to
strengthen mass segregation trends, and that the strength of mass
segregation tends to decrease with increasing group halo mass. We find
the same trends if we use the fraction of massive galaxies as a
function of group-centric radius as an alternative probe of mass
segregation.  The magnitude of mass
segregation that we measure, particularly in high-mass haloes,
indicates that dynamical friction is not acting efficiently.
\end{abstract}

\begin{keywords}
galaxies: clusters: general -- galaxies: evolution -- galaxies:
groups: -- galaxies: statistics
\end{keywords}

\section{Introduction}
\label{sec:introduction}
It has been well established that galaxy properties depend strongly
on local environment \citep[e.g.][]{oemler1974, hogg2004,
  blanton2005b, tal2014}. Galaxies in dense environments such as
clusters tend to have lower star formation rates (SFRs), while isolated field
galaxies are generally actively forming stars
\citep[e.g.][]{balogh2000, ball2008, wetzel2012}.  It is also well known
that galaxy properties, like SFR, depend strongly on galaxy mass
\citep[e.g.][]{poggianti2008}.  It is critical to study the
distribution of galaxy masses within haloes of different masses in
order to ascertain whether the variations in galaxy properties with
environment are due to physical
mechanisms acting in dense environments, or simply due to the fact that
high density environments contain more high mass galaxies.
Intermediate density environments, galaxy groups,
represent not only the most common environment in the local universe
\citep{geller1983, eke2005}, but also represent the environment where
many physical processes are efficient.  Galaxy interactions like
mergers and harassment are favoured in this environment because of the
low relative velocities between galaxies \citep{zabludoff1998, brough2006}.
\par
The study of mass segregation in groups can be used to elucidate
information on physical processes such as dynamical friction, galaxy
mergers, and tidal stripping. Mass segregation in bound structures has
generally been predicted as a result of
dynamical friction \citep{chandrasekhar1943}. Dynamical friction
acts as a drag force on orbiting bodies and massive
galaxies within groups and clusters are expected to migrate to
smaller radii
as time progresses. If dynamical friction is a dominant factor then
clear mass segregation should be observable in evolved groups and
clusters.
\par
Galaxy groups are not static systems, but are constantly being
replenished by infalling galaxies from the field. Infalling galaxies
are preferentially found at large radii \citep{wetzel2013} and the
difference in stellar mass distributions between evolved group members
and infalling galaxies could affect the strength of mass segregation.
\par
If significant mass segregation is not found, then this
implies that either: the timescale associated with dynamical friction
is greater than the age of the group/cluster, or that there are other
physical processes, such as merging, tidal stripping, or
pre-processing, which are
playing a more important role than dynamical friction.
\par
Recent work has shown conflicting results with regards to the presence
of mass segregation in groups and clusters. \citet{ziparo2013} find
no evidence for strong mass segregation in X-ray
selected groups from the ECDFS, COSMOS, GOODS-North, and GOODS-South
fields out to $z = 1.6$, for a sample of galaxies with
$M_{\mathrm{star}} >
10^{10.3}\,\mathrm{M_{\odot}}$. \citet{vonderlinden2010} examine SDSS 
galaxy clusters and find no
evidence for mass segregation in four different redshift bins at $z <
0.1$.  von der Linden et al. make redshift dependent stellar mass cuts
ranging from $10^{9.6}\,\mathrm{M_{\odot}}$ to
$10^{10.5}\,\mathrm{M_{\odot}}$. \citet{vulcani2013}
use mass limited samples at $0.3 \le z \le 0.8$ from the IMACS Cluster
Building Survey and
the ESO Distant Cluster Survey, with stellar mass cuts at
$M_{\mathrm{star}} > 10^{10.5}\,\mathrm{M_{\odot}}$ and $M_{\mathrm{star}} >
10^{10.2}\,\mathrm{M_{\odot}}$ respectively, to study galaxy stellar mass
functions in different environments. Vulcani et al. find no
statistical differences between mass functions of galaxies located at
different cluster-centric distances.
\par
Conversely, \citet{balogh2014} find evidence for mass segregation
in GEEC2 groups at $0.8 < z < 1$, using a stellar mass limited sample
with $M_{\mathrm{star}} > 10^{10.3}\,\mathrm{M_{\odot}}$. Using a
volume limited sample of zCOSMOS groups \citet{presotto2012}
find evidence for mass segregation in their whole sample at both $0.2
\le z \le 0.45$ and $0.45 \le z \le 0.8$. Presotto et al. also
break their sample into rich and poor groups at $0.2 \le z \le 0.45$,
and find evidence for mass segregation within rich groups but find no
evidence for mass segregation within poor groups. Using a
$V_{\mathrm{max}}$ weighted sample with a stellar mass cut at
$10^{9.0}\,\mathrm{M_{\odot}}$,
\citet{vandenbosch2008} find evidence for mass segregation in SDSS
groups. 
\par
It is clear that there lacks consensus with regards to the
strength of mass 
segregation, or its halo mass dependence. 
\par
In this letter we present evidence of the presence of a small, but
significant, amount
of mass segregation in SDSS galaxy groups. We
show that the detection of mass segregation is dependent on stellar
mass completeness, with completeness cuts at relatively high stellar
masses potentially masking underlying mass segregation trends. We also
show that the strength of mass
segregation scales inversely with halo mass, with cluster
sized haloes showing little to no observable mass segregation. In
\S~\ref{sec:data} we briefly describe our data set, in
\S~\ref{sec:results} we present our results from this work, in
\S~\ref{sec:discussion} we provide a discussion of our results,
and in \S~\ref{sec:conclusion} we give a summary of the results and make concluding statements.
\par
In this letter we assume a flat $\Lambda$CDM cosmology with $\Omega_\mathrm{M} =
0.3$, $\Omega_\Lambda = 0.7$, and $H_0 = 70\,\mathrm{km\,s^{-1}\,
  Mpc^{-1}}$.


\section{Data}
\label{sec:data}

The results presented in this letter utilize the group catalogue of
\citet{yang2007}. This catalogue is constructed by applying the
halo-based group finder of \citet{yang2005, yang2007} to the New York
University Value-Added
Galaxy Catalogue (NYU-VAGC; \citealt{blanton2005}), which is based on the Sloan
Digital Sky Survey Data Release 7 (SDSS DR7;
\citealt{abazajian2009}). Stellar masses are obtained from the
NYU-VAGC and are computed using
the methodology of \citet{blanton2007}, assuming a
\citet{chabrier2003} initial mass function. Halo
masses are determined using the ranking of the characteristic stellar
mass, $M_{\mathrm{*,\,grp}}$, and assuming a relationship between
$M_{\mathrm{halo}}$ and $M_{\mathrm{*,\,grp}}$
\citep{yang2007}.  $M_{\mathrm{*,\,grp}}$ is
  defined by Yang et al. as 

\begin{equation}
  M_{\mathrm{*,\,grp}} =
  \frac{1}{g(L_{19.5},\,L_{\mathrm{lim}})}\sum_i\frac{M_{\mathrm{star},\,i}}{C_i},
\end{equation}

\noindent
where $M_{\mathrm{star},\,i}$ is the stellar mass of the $i^{\mathrm{th}}$ member
galaxy, $C_i$ is the completeness of the survey at the position of
that galaxy, and $g(L_{19.5},\,L_{\mathrm{lim}})$ is a correction
factor which accounts for galaxies missed due to the magnitude limit
of the survey.
\par
Halo-centric distance for each galaxy is not given
explicitly in the Yang catalog, however we calculate it using the
redshift of the group and the angular separation of the galaxy and
halo centre on the sky. We measure group-centric radius from the
luminosity weighted centre of each group, and normalize our
group-centric radii by $R_{200}$. We use the definition for $R_{200}$
as
given in \citet{carlberg1997}

\begin{equation}
  R_{200} = \frac{\sqrt{3}\sigma}{10H(z)},
\end{equation}

\noindent
where the Hubble parameter, $H(z)$, is defined as

\begin{equation}
  H(z) = H_0\sqrt{\Omega_\mathrm{M}(1+z)^3 + \Omega_\Lambda},
\end{equation}

\noindent
and we calculate the velocity dispersion, $\sigma$, as 

\begin{equation}
  \sigma =
  397.9\,\mathrm{km}\,\mathrm{s^{-1}}\left(\frac{M_{\mathrm{halo}}}{10^{14}\,h^{-1}\,\mathrm{M_{\odot}}}\right)^{0.3214},
\end{equation}

\noindent
where the above is a fitting function given in \citet{yang2007}.
\par
For our analysis we select group galaxies with redshift, $z < 0.1$, that are
within two virial radii of the group centre, and
groups with a minimum of three galaxy members -- although our results
are not sensitive to these specific cuts. For our
sample over 95 per cent of group galaxies reside within two virial
radii of the group centre. We
also subtract the
most massive galaxy (MMG) from each group, to ensure that any
underlying radial mass trend is not contaminated by the MMG.
\par
This sample is not volume limited, therefore the sample will suffer
from Malmquist bias. This leads to a bias towards objects of higher
luminosity and stellar mass, with increasing redshift. To
account for this bias we weight our sample by $1/V_{\mathrm{max}}$,
where $V_{\mathrm{max}}$ is the comoving volume of the universe out to
a comoving radius at which the galaxy would have met the selection
criteria for the sample. For our $V_{\mathrm{max}}$ weights we apply the
values presented in the catalogue of \citet{simard2011} to our sample.
\par
In order to investigate the effect of stellar mass limits on the
detection of mass
segregation, we use samples corresponding to various stellar mass
cuts. We perform our analysis on an unweighted
sample with two mass cuts corresponding to $M_{\mathrm{star}} >
10^{10.5}\,\mathrm{M_{\odot}}$ (4152 galaxies in 1970 groups) and $M_{\mathrm{star}} >
10^{10.0}\,\mathrm{M_{\odot}}$ (26774 galaxies in 4534 groups), and a
$V_{\mathrm{max}}$ weighted sample
with mass cuts at $M_{\mathrm{star}} > 10^{9.0}\,\mathrm{M_{\odot}}$
(56957 galaxies in
7217 groups) and $M_{\mathrm{star}} >
10^{8.5}\,\mathrm{M_{\odot}}$ (59791 galaxies in 7289 groups). The
unweighted sample
is stellar mass complete down to $M_{\mathrm{star}} >
10^{10.0}\,\mathrm{M_{\odot}}$. Therefore, for both the weighted and
unweighted sample, we have two different stellar mass cuts, giving us four
separate samples in total.

\begin{figure*}
  \centering
  \includegraphics[width = \textwidth]{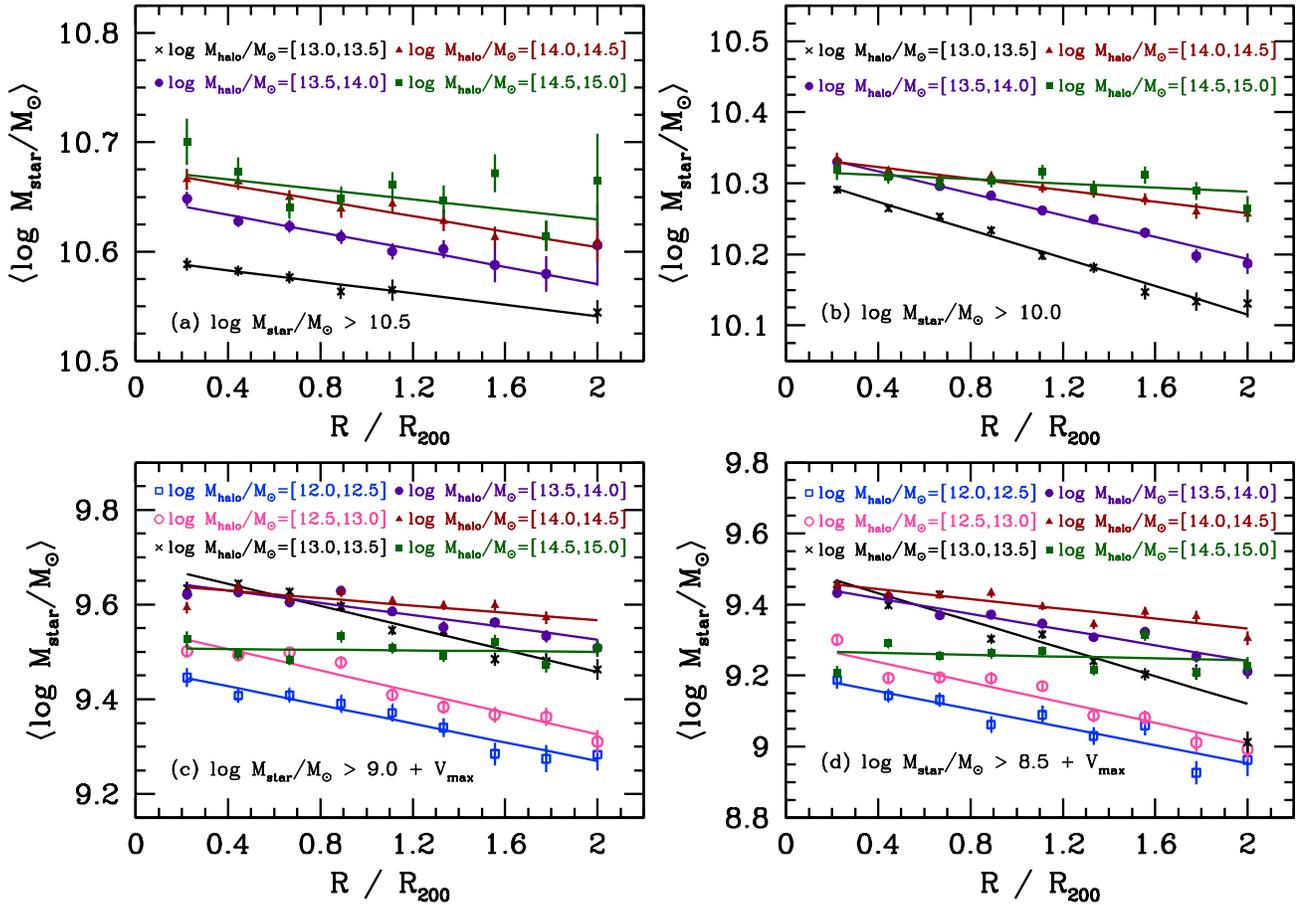}
  \caption{All panels show mean mass as a function of normalized radial
    distance for various halo mass bins, with error bars corresponding
    to $1\sigma$ statistical errors. The solid
      lines correspond to weighted least-squares fits for each halo
      mass bin. \emph{Top left:} Unweighted sample,
    for galaxies with $\mathrm{log}(M_{\mathrm{star}} /
    \mathrm{M_{\odot}}) > 10.5$. \emph{Top right:} Unweighted sample,
    for galaxies with $\mathrm{log}(M_{\mathrm{star}} / \mathrm{M_{\odot}}) >
    10.0$.  \emph{Botton left:} $V_{\mathrm{max}}$ weighted sample,
    for galaxies with $\mathrm{log}(M_{\mathrm{star}} /
    \mathrm{M_{\odot}}) > 9.0$. \emph{Bottom Right:}
    $V_{\mathrm{max}}$ weighted sample,
    for galaxies with $\mathrm{log}(M_{\mathrm{star}} /
    \mathrm{M_{\odot}}) > 8.5$. Note that
  different mass scales are used in each panel. There are more halo
  mass bins in the bottom row due to the increased number of
  low mass galaxies as a result of $V_{\mathrm{max}}$ weighting.}
  \label{fig:RmassH}
\end{figure*}


\section{Results}
\label{sec:results}

\subsection{Mass segregation in SDSS groups}
\label{sec:massSeg}

In Fig.~\ref{fig:RmassH} we plot mean stellar mass as a function of
radial distance from the group centre for various halo mass
bins. Fig.~\ref{fig:RmassH}a corresponds to our high-mass cut, unweighted
sample, Fig.~\ref{fig:RmassH}b corresponds to our low-mass cut,
unweighted sample, Fig.~\ref{fig:RmassH}c corresponds to our
high-mass cut, weighted sample, and Fig.~\ref{fig:RmassH}d corresponds to our
low-mass cut, weighted sample.
\par
For all halo mass
bins, and regardless of mass cut, the unweighted sample shows
statistically significant mass segregation with a weighted linear
least-squares fit. The $V_{\mathrm{max}}$
weighted sample shows statistically significant mass segregation for the five lower halo mass bins,
whereas the highest halo mass bin has a best-fitting slope consistent with
zero -- this trend holds for both mass cuts.  For both the weighted
and unweighted samples there is a clear trend of the slope with
halo mass -- more massive haloes show weaker mass segregation.  This
result will be discussed in \S~\ref{sec:haloMass}.
\par
We find that our highest halo mass sample ($M_{\mathrm{halo}} >
10^{14.5}\,M_{\odot}$) has a large number of low mass galaxies when
compared to the high halo mass samples, which leads to a smaller mean
stellar mass in the $V_{\mathrm{max}}$ weighted results shown in
Fig.~\ref{fig:RmassH}c \& \ref{fig:RmassH}d.  While this introduces a
shift in normalization, it does not affect the mass segregation trend
and therefore does not change the key result that mass segregation
depends on halo mass.

\subsection{Massive galaxy fraction}
\label{sec:massiveGal}

\begin{figure*}
  \centering
  \includegraphics[width = \textwidth]{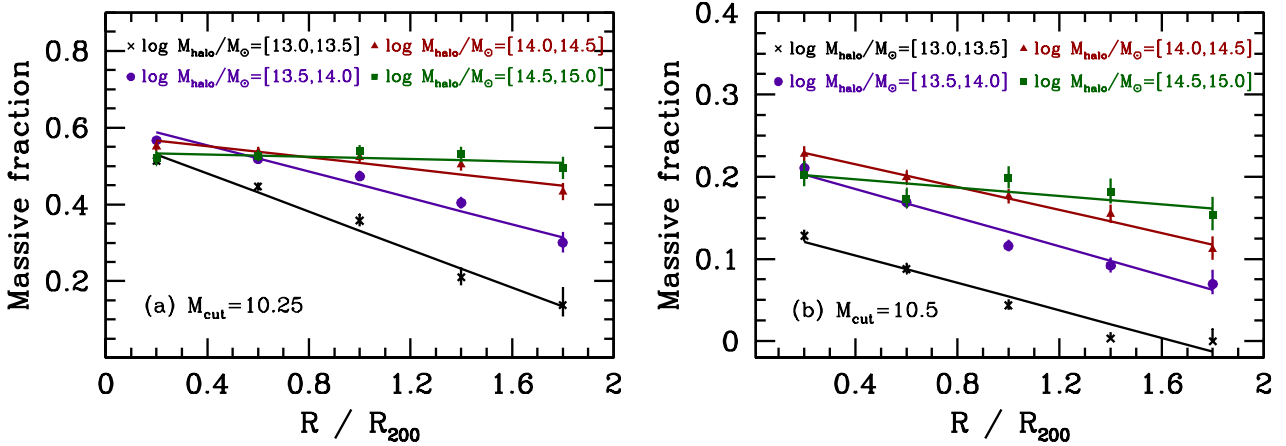}
  \caption{Fraction of massive galaxies with respect to normalized radial
    distance. Error bars are given by a
    $1\sigma$ binomial confidence
    interval, calculated using the beta distribution as outlined in
    \citet{cameron2011}.  The solid lines
      correspond to weighted least-squares fits for each halo mass
      bin. \emph{Left panel:} The fraction of
    galaxies with $\mathrm{log}(M_{\mathrm{star}} /
    \mathrm{M_{\odot}}) > 10.25$ as a function of radial distance, for
  the unweighted sample with $M_{\mathrm{star}} >
  10^{10}\,\mathrm{M_{\odot}}$. \emph {Right
      panel:} The fraction of
    galaxies with $\mathrm{log}(M_{\mathrm{star}} /
    \mathrm{M_{\odot}}) > 10.5$ as a function of radial distance, for
  the unweighted sample with $M_{\mathrm{star}} >
  10^{10}\,\mathrm{M_{\odot}}$.}
  \label{fig:RfracH}
\end{figure*}

An alternative way to investigate galaxy populations within the group
sample is to study the fraction of `massive' galaxies at various
group-centric radii. In Fig.~\ref{fig:RfracH} we plot the fraction of
massive galaxies as a function of radial distance. We
calculate the massive fraction for each radial bin as

\begin{equation}
  f_m(M_{\mathrm{cut}}) = \frac{\mathrm{\#\;galaxies\;with}\;M_{\mathrm{star}} >
    M_{\mathrm{cut}}}{\mathrm{\#\;galaxies\;with}\;M_{\mathrm{star}}
  > 10^{10}\,\mathrm{M_{\odot}}},
\end{equation}

\noindent
where $M_{\mathrm{cut}}$ is a stellar mass cut-off above which we
define a massive galaxy. We initially apply a high mass galaxy cut,
$M_{\mathrm{cut}}$, at $10^{10.25}\,\mathrm{M_{\odot}}$ corresponding
to the median stellar mass of the unweighted sample (with the low-mass
cut at $10^{10}\,\mathrm{M_{\odot}}$). Comparing Fig.~\ref{fig:RfracH}a and
Fig.~\ref{fig:RmassH}b we see essentially identical
trends. We observe the same trends of mass segregation whether
we look at the average galaxy mass at a given radius, or consider the
fraction of massive galaxies.
\par
To confirm that this trend is robust regardless of the mass cut-off
used to define a massive galaxy, we make the
same plot but now use $M_{\mathrm{cut}} = 10^{10.5}\,\mathrm{M_{\odot}}$.
Comparing Fig.~\ref{fig:RfracH}a and \ref{fig:RfracH}b we see
that while the overall fractions of massive galaxies decrease with
increasing the stellar mass cut, the
trend essentially stays
the same. There is clear evidence for mass segregation and the
strength of the mass segregation depends on halo mass.


\section{Discussion}
\label{sec:discussion}

\subsection{Effect of including low mass galaxies}
\label{sec:lowMass}

The results in Fig.~\ref{fig:RmassH} show that mass segregation
generally increases when lower mass
galaxies are included. To quantify this
effect we can compare the best-fitting slopes corresponding to the
high-mass and the low-mass cut samples.
\par  
For a given halo mass, the low-mass cut sample
displays larger slopes than the high-mass cut sample for two of the
halo mass bins. The slopes corresponding to the other two halo mass
bins are consistent with being equal. For the weighted sample we find
similar results with the low-mass cut sample showing larger slopes for
three of the halo mass bins, and the other three halo mass bins
showing slopes consistent with being equal.
\par
This suggests that the inclusion of low-mass galaxies has a measurable
effect on the observation of mass
segregation. Studies which make mass cuts at moderate to
high stellar mass, are potentially missing a mass segregation contribution
from low-mass galaxies.  We note that these results are somewhat in
disagreement with \citet{ziparo2013} who find no mass segregation in
their sample even with the inclusion of low-mass galaxies.  The mass
segregation we observe in Fig.~\ref{fig:RmassH} \& \ref{fig:RfracH} is
very weak and the sample of Ziparo et al. may have been
too small to show this subtle trend.

\subsection{Halo mass dependence}
\label{sec:haloMass}

Fig.~\ref{fig:RmassH} \& \ref{fig:RfracH} clearly
indicate that the highest halo mass bins show the least mass
segregation. This trend is consistent in all cases, regardless of
stellar mass cut or whether the sample had $V_{\mathrm{max}}$ weights
applied. Our observed dependence on halo mass is consistent with
results finding no measurable mass
segregation in galaxy clusters \citep{pracy2005, vonderlinden2010,
  vulcani2013}.
\par
It has been shown through N-body simulations that the dynamical friction
timescale scales
with $M_{\mathrm{h}}/M_{\mathrm{s}}$ \citep[e.g.][]{taffoni2003,
  conroy2007, boylan2008},
where $M_{\mathrm{s}}$ is the
initial satellite mass and $M_{\mathrm{h}}$ is the mass of the host
halo. For a given
satellite mass, this implies a longer dynamical friction timescale
for larger haloes, which is consistent with our result. This can be
interpretted as an increase in tidal stripping efficiency as
$M_{\mathrm{h}}/M_{\mathrm{s}}$ increases
\citep{taffoni2003}. \citet{gan2010} have shown
that for an infalling satellite the dynamical friction timescale
increases with a stronger tidal field.  This is due to tidal
stripping retarding the decay of satellite angular momentum, which
increases the dynamical friction timescale.
\par
It should also be noted that the merger timescale scales with
$M_{\mathrm{s}}/M_{\mathrm{h}}$ \citep{jiang2008}, which implies a higher merger
efficiency in low mass haloes, for a given satellite mass. The
build-up of massive objects through galaxy mergers could enhance mass
segregation in low-mass haloes, in accordance with our results.
\par
There has been evidence of cluster galaxies having their star
formation quenched in lower mass groups ($\sim10^{13}\,\mathrm{M_{\odot}}$)
prior to accretion into the cluster environment
\citep[e.g.][]{zabludoff1998, mcgee2009, delucia2012, hou2014}. This
pre-processing could potentially provide an explanation of our
observed mass segregation trends with halo mass. If mass segregation
is present in the group environment as a result of pre-processing, the
recent accretion of multiple pre-processed groups to form a galaxy cluster
could result in little to no observed mass segregation in the cluster
as a whole.
In other words, if the cluster environment consists of multiple subhaloes
at various cluster-centric radii, while individual subhaloes may show
mass segregation, the total effect of these subhaloes together may
leave the cluster with a relatively uniform radial mass distribution.
\par
\citet{vulcani2014} apply semi-analytic models to the Millenium
Simulation \citep{springel2005} to study galaxy mass functions in
different environments.
Vulcani et al. simulate galaxy mass functions for three halo masses,
$\log (M_{\mathrm{halo}}/\mathrm{M_{\odot}})=\{13.4,\,14.1,\,15.1\}$,
as a function of cluster-centric radius. In the lowest mass halo they
find the mass function depends slightly on cluster-centric radius,
with the innermost regions showing flatter mass functions at low and
intermediate masses. This trend persists, but is not as strong at
intermediate halo mass. The highest halo mass bin shows virtually
identical mass function shapes for all cluster-centric radii. This
result is indicative of measurable mass segregation for the low
and intermediate mass haloes, with the strength of mass segregation
decreasing with increasing halo mass. These simulation trends show
excellent agreement with our observed dependence of mass segregation
on halo mass.

\subsection{Reconciling previous results}
\label{prevResults}

In \S~\ref{sec:introduction} we mention previous literature results
which present evidence both for and against the presence of mass
segregation in groups and clusters. We argue that the majority of
these results can be reconciled with our two main findings:

\begin{enumerate}
  \item Mass segregation is enhanced with the inclusion of
    low-mass galaxies in a sample.
  \item Mass segregation decreases with increasing halo mass,
    with high-mass haloes showing little to no mass segregation.
\end{enumerate}

\noindent
Of the studies mentioned in \S~\ref{sec:introduction}, those which
observe no evidence for mass segregation either: make a mass completeness cut at
intermediate to high stellar mass, or observe this lack of mass
segregation only in high-mass haloes. Therefore the lack of observed
mass segregation can potentially be explained through the lack of
low-mass galaxies in the study survey, or the study being limited to
high halo mass environments.


\section{Conclusion}
\label{sec:conclusion}

In this letter we examine mass segregation trends in the
\citet{yang2007} SDSS DR7 groups
for various stellar and halo mass cuts. We show that a small, but
significant, amount of mass segregation is
present in these groups. This mass segregation shows consistent
trends, with lower stellar mass samples showing stronger mass
segregation, and groups residing in large haloes showing little to no
mass segregation.
\par
The magnitude of mass segregation we measure, especially in high mass
haloes, is potentially indicative of dynamical friction not acting
efficiently.  We discuss previous
literature to provide possible explanations for the observed
trends, showing that our observed trends with halo mass agree with
prior results. Further work with hydrodynamic simulations would
be helpful 
to further constrain the important mechanisms responsible for our
observed mass trends and the lack of mass segregation in high-mass haloes. 

\section*{Acknowledgments}
\label{sec:acknowledgments}

We thank the anonymous referee for their various helpful comments and
suggestions.  IDR and LCP thank the National Science and Engineering Research
Council of Canada for funding.  We thank X. Yang et al. for
making their
SDSS DR7 group catalogue public, L. Simard et al. for the
publication of their SDSS DR7 morphology catalogue, and the NYU-VAGC
team for the 
publication of their SDSS DR7 catalogue.  This research would not have
been possible without these public catalogues.
\par
Funding for the SDSS has been provided by the Alfred P. Sloan
Foundation, the Participating Institutions, the National Science
Foundation, the U.S. Department of Energy, the National Aeronautics
and Space Administration, the Japanese Monbukagakusho, the Max Planck
Society, and the Higher Education Funding Council for England. The
SDSS Web Site is http://www.sdss.org/.


\begin{thebibliography}{}
\bibitem[\protect\citeauthoryear{Abazajian et
    al.}{2009}]{abazajian2009} Abazajian K.N. et al., 2009,
ApJS, 182, 543

\bibitem[\protect\citeauthoryear{Ball, Loveday \&
    Brunner}{2008}]{ball2008} Ball N.M., Loveday J, Brunner R.J., 2008,
MNRAS, 383, 907

\bibitem[\protect\citeauthoryear{Balogh et
    al.}{2014}]{balogh2014} Balogh M.L. et al., 2014,
MNRAS, 443, 2679

\bibitem[\protect\citeauthoryear{Balogh, Navarro \&
    Morris}{2000}]{balogh2000} Balogh M.L., Navarro J.F., Morris S.L., 2000,
ApJ, 540, 113

\bibitem[\protect\citeauthoryear{Blanton \&
    Roweis}{2007}]{blanton2007} Blanton M.R., Roweis S., 2007,
AJ, 133, 734

\bibitem[\protect\citeauthoryear{Blanton et
    al.}{2005}]{blanton2005b} Blanton M.R., Eisenstein D., Hogg D.W.,
  Schlegel D.J., Brinkmann J., 2005,
ApJ, 629, 143

\bibitem[\protect\citeauthoryear{Blanton et
    al.}{2005}]{blanton2005} Blanton M.R. et al., 2005,
ApJ, 129, 2562

\bibitem[\protect\citeauthoryear{Boylan-Kolchin, Ma \&
    Quataert}{2008}]{boylan2008} Boylan-Kolchin M., Ma C.P., Quataert E., 2008,
MNRAS, 383, 93

\bibitem[\protect\citeauthoryear{Brough et al.}{2006}]{brough2006}
  Brough S., Forbes D.A., Kilborn V.A., Couch W., 2006, MNRAS, 370, 1223

\bibitem[\protect\citeauthoryear{Cameron}{2011}]{cameron2011} Cameron E., 2011,
PASA, 28, 128

\bibitem[\protect\citeauthoryear{Carlberg et
    al.}{1997}]{carlberg1997} Carlberg R.G. et al., 1997,
ApJ, 485, L13

\bibitem[\protect\citeauthoryear{Chabrier}{2003}]{chabrier2003}
  Chabrier G., 2003,
PASP, 115, 763

\bibitem[\protect\citeauthoryear{Chandrasekhar}{1943}]{chandrasekhar1943}
  Chandrasekhar S., 1943,
ApJ, 97, 255

\bibitem[\protect\citeauthoryear{Conroy, Ho \&
    White}{2007}]{conroy2007} Conroy C., Ho S., White M., 2007,
MNRAS, 379, 1491

\bibitem[\protect\citeauthoryear{De Lucia et
    al.}{2012}]{delucia2012} De Lucia G., Weinmann S., Poggianti B.M.,
  Arag{\'o}n-Salamanca A., Zaritsky D., 2012,
MNRAS, 423, 1277

\bibitem[\protect\citeauthoryear{Eke et
    al.}{2005}]{eke2005} Eke V.R., Baugh C.M., Cole S., Frenk C.S.,
  King H.M., Peacock J.A., 2005, MNRAS, 362, 1233

\bibitem[\protect\citeauthoryear{Gan et
    al.}{2010}]{gan2010} Gan J.L., Kang X., Hou J.L., Chang R.X., 2010,
MNRAS, 10, 1242

\bibitem[\protect\citeauthoryear{Geller \& Huchra}{1983}]{geller1983}
  Geller M.J., Huchra J.P., 2010, ApJS, 52, 61

\bibitem[\protect\citeauthoryear{Hogg et
    al.}{2004}]{hogg2004} Hogg D.W. et al., 2004, ApJ, 601, L29

\bibitem[\protect\citeauthoryear{Hou, Parker \&
    Harris}{2014}]{hou2014} Hou A., Parker L.C., Harris W.E., 2014,
MNRAS, 442, 406

\bibitem[\protect\citeauthoryear{Jiang et
    al.}{2008}]{jiang2008} Jiang C.Y., Jing Y.P., Faltenbacher A.,
  Lin W.P., Lin C., 2008, ApJ, 675, 1095

\bibitem[\protect\citeauthoryear{Maccio et
    al.}{2007}]{maccio2007} Macci{\`o} A.V., Dutton A.A., van den
  Bosch F.C., Moore B., Potter D., Stadel J., 2007, MNRAS, 378, 55

\bibitem[\protect\citeauthoryear{McGee et
    al.}{2009}]{mcgee2009} McGee S.L., Balogh M.L., Bower R.G., Font
  A.S., McCarthy I.G., 2009, MNRAS, 400, 937

\bibitem[\protect\citeauthoryear{Oemler}{1974}]{oemler1974} Oemler
  J. A., 1974, ApJ, 194, 1

\bibitem[\protect\citeauthoryear{Poggianti et
    al.}{2008}]{poggianti2008} Poggianti B.M. et al., 2008, ApJ, 684, 888

\bibitem[\protect\citeauthoryear{Pracy et
    al.}{2005}]{pracy2005} Pracy M.B., Driver S.P., De Propris R.,
  Couch W.J., Nulsen P.E.J., 2005, MNRAS, 364, 1147

\bibitem[\protect\citeauthoryear{Presotto et
    al.}{2012}]{presotto2012} Presotto et al., 2012, A\&A, 539, A55

\bibitem[\protect\citeauthoryear{Simard et
    al.}{2011}]{simard2011} Simard L., Mendel J.T., Patton D.R.,
  Ellison S.L., McConnachie A.W., 2011, ApJS, 196, 11

\bibitem[\protect\citeauthoryear{Springel et al.}{2005}]{springel2005}
  Springel V. et al., 2005, Nature, 435, 629 

\bibitem[\protect\citeauthoryear{Taffoni et
    al.}{2003}]{taffoni2003} Taffoni G., Mayer L, Colpi M., Governato
  F., 2003, MNRAS, 341, 434

\bibitem[\protect\citeauthoryear{Tal et
    al.}{2014}]{tal2014} Tal T. et al., 2014, ApJ, 789, 164

\bibitem[\protect\citeauthoryear{van den Bosch et
    al.}{2004}]{vandenbosch2004} van den Bosch F.C., Norberg P., Mo H.J., Yang
  X., 2004, 352, 1302

\bibitem[\protect\citeauthoryear{van den Bosch et
    al.}{2008}]{vandenbosch2008} van den Bosch F.C., Pasquali A., Yang
  X., Mo H.J., Weinmann S., McIntosh D.H., Aquino D., 2008, preprint
  (arXiv0805.0002)

\bibitem[\protect\citeauthoryear{von der Linden et
    al.}{2010}]{vonderlinden2010} von der Linden A., Wild V.,
  Kauffmann G., White S.D.M., Weinmann S., 2010, MNRAS, 404, 1231

\bibitem[\protect\citeauthoryear{Vulcani et
    al.}{2013}]{vulcani2013} Vulcani B. et al., 2013, A\&A, 550, A58

\bibitem[\protect\citeauthoryear{Vulcani et
    al.}{2014}]{vulcani2014} Vulcani B., De Lucia G., Poggianti B.M.,
  Bundy K., More S., Calvi R., 2014, ApJ, 788, 57

\bibitem[\protect\citeauthoryear{Wetzel, Tinker \&
    Conroy}{2012}]{wetzel2012} Wetzel A.R., Tinker J.L., Conroy C.,
  2012, MNRAS, 424, 232

\bibitem[\protect\citeauthoryear{Wetzel et al.}{2013}]{wetzel2013}
  Wetzel A.R., Tinker J.L., Conroy C., van den Bosch F.C.,
  2013, MNRAS, 432, 336

\bibitem[\protect\citeauthoryear{Yang et
    al.}{2005}]{yang2005} Yang X., Mo H.J., van den Bosch F.C., Jing
  Y.P., 2005, MNRAS, 356, 1293

\bibitem[\protect\citeauthoryear{Yang et
    al.}{2007}]{yang2007} Yang X., Mo H.J., van den Bosch F.C.,
  Pasquali A., Li C., Barden M., 2007, ApJ, 671, 153

\bibitem[\protect\citeauthoryear{Zabludoff \&
    Mulchaey}{1998}]{zabludoff1998} Zabludoff A.I., Mulchaey J.S.,
  1998, ApJ, 496, 39

\bibitem[\protect\citeauthoryear{Ziparo et
    al.}{2013}]{ziparo2013} Ziparo F. et al., 2013, MNRAS, 434, 3089

\end{thebibliography}
\end{document}